\documentclass{article}
\usepackage[english]{babel}
\usepackage{authblk}
\usepackage{ulem}
\usepackage{amsmath}
\usepackage{blindtext}
\usepackage{esvect}
\usepackage{bbold}
\usepackage{bm}
\usepackage{comment}

\usepackage[letterpaper,top=2cm,bottom=2cm,left=3cm,right=3cm,marginparwidth=1.75cm]{geometry}

\usepackage{amsmath}
\usepackage{graphicx}
\usepackage[colorlinks=true, allcolors=blue]{hyperref}

\title{Verification of conditional mechanical squeezing for a mg-scale pendulum near quantum regimes}

\author[a]{Jordy G. Santiago-Condori}
\author[b]{Naoki Yamamoto}
\author[c]{Nobuyuki Matsumoto} 
\affil[a]
{Research Institute of Electrical Communication, Tohoku University, Sendai 980-8577, Japan}
\affil[b]{Department of Applied Physics and Physico-Informatics, Keio University, Hiyoshi 3-14-1, Kohoku, Yokohama 223- 8522, Japan}
\affil[c]{Department of Physics, Gakushuin University, Mejiro 1-5-1, Toshima, Tokyo 171-8588, Japan}
\newcommand{\simgt}{\lower.5ex\hbox{$\; \buildrel > \over \sim \;$}}
\newcommand{\simlt}{\lower.5ex\hbox{$\; \buildrel < \over \sim \;$}}

\begin{document}
\maketitle
{ E-mail: matsumoto.granite@gmail.com}

\begin{abstract}
In quantum mechanics, measurement can be used to prepare a quantum state. 
This principle is applicable even for macroscopic objects, which may enable us to see classical-quantum transition. 
Here, we demonstrate conditional mechanical squeezing of a mg-scale suspended mirror (i.e. the center-of-mass mode of a pendulum) near quantum regimes, through continuous linear position measurement and quantum state prediction. 
The experiment involved the pendulum interacting with photon coherent fields in a detuned optical cavity, which creates an optical spring. 
Futhermore, the detuned cavity allows us to perform linear position measurement by direct photo-detection of the reflected light. 
We experimentally verify the conditional squeezing using the theory combining prediction and retrodiction based on the causal and anti-causal filters. 
As a result, the standard deviation of position and momentum are respectively given by 36 times the zero-point amplitude of position $q_{\rm zpf}$ and 89 times the zero-point amplitude of momentum $p_{\rm zpf}$. 
The squeezing level achieved is about 5 times closer to the zero-point motion, despite that the mass of the mechanical oscillator is approximately 7 orders of magnitude greater, compared to the previous study. 
Thus, our demonstration is the first step towards quantum control for massive objects whose mass-scale is high enough to measure gravitational interactions. 
Such quantum control will pave the way to test quantum mechanics using the center-of-mass mode of massive objects. 
\end{abstract}

\section*{Introduction}
The investigation of continuous linear position measurements of macroscopic objects has been mainly motivated by the direct detection of gravitational waves \cite{ligo1,ligo2}, and the field of cavity optomechanics \cite{aspelmeyer}. 
These research established the standard quantum limit (SQL) for continuous position measurements \cite{SQL0,SQL1}, where shot noise and quantum back-action noise contribute equally. 
Such precise measurements allow measurement-based quantum control of macroscopic objects, like  ground state cooling \cite{ground1,ground2,ground3} and generation of entanglement \cite{entangle2,macroscopic}, since correlations are built up between the mechanical objects and the  measuring devices via radiation pressure of light. 
It is expected that, as the measurement sensitivity increases for example by enhancing the mechanical quality factor \cite{Q0,Q1,Q2,Q3}, tests with quantum oscillators of unexplored phenomena such as gravity decoherence \cite{penrose,diosi,bassi,GD,kanno1,kanno2}, semiclassical gravity \cite{semiG0,semiG1,semiG2}, and dark matter-induced fifth force \cite{fifth0,fifth1,fifth2} will become possible.  
One of the most challenging tasks is to measure entanglement via Newtonian gravity in order to test the quantum nature of the gravitational interaction \cite{QG1,QG2,QG3,QG4,QG5}. 
Towards this ultimate end, it is important to first develop technology to control the center-of-mass mode of massive objects via laser light. 

Recently, Meng {\it et al.} \cite{MS0, multimode} demonstrated theoretically and experimentally that mechanical squeezing can be generated outside the rotating wave approximation regime, by applying a causal Wiener filter \cite{wiener}, which minimizes the mean-square estimation error for position monitoring. 
While in \cite{MS0} Meng {\it et al.} consider an optomechanical system consisting of an optical cavity on resonance coupled to a mechanical oscillator, in our analysis we take into account a detuning from resonance. 
This is because a detuned cavity allows us to create an optical spring \cite{OS0,Q0,OS1} that optically traps a massive mechanical oscillator of low rigidity (e.g. the pendulum mode of a suspended mirror). %% which might be difficult without optical spring-->Suggest removing this SBCL. 
Importantly, this lets us sufficiently increase the quantum coherence time of the mechanical mode \cite{OS2}.  
Furthermore, the detuned cavity allows us to monitor the position via direct photo-detection, which is the simplest configuration among various optical detection schemes, such as homodyne detection. 
Because generation of entanglement normally requires the use of a complex Power-Recycling Fabry-Perot Michelson Interferometer (PRFPMI) \cite{entangle2, macroscopic}, simplifying the detection system is extremely important to increase the feasibility of the experiment.

In this paper, we derive the analytic solution of a causal Wiener filter for preparing conditional squeezing as well as an anti-causal Wiener filter for verifying the state. 
Since conditional states are characterized by conditional variances (the mean of the squared difference between the {\it true} value and the estimated value by the causal filter), it requires us to obtain experimental access of the true value for determining the state. 
This {\it true} value, on the other hand, is only available mathematically, especially in case where the system is macroscopic. 
To avoid this difficulty, we experimentally verify conditional variances by comparing the results of causal and anti-causal estimation following Rossi {\it et al} \cite{verification}. 
According to this process, the conditional state can be verified only by using the measured data and the filters including system parameters, independent of quantum physical properties such as position and momentum. 

Here, we present experimental verification for the conditional squeezing in the center-of-mass of an optically trapped mg-scale pendulum (resonance at 280 Hz) near quantum regimes, by applying causal and anti-causal Wiener filters to the position measurement record previously reported for gravity sensing \cite{miliG1}.
The verified position (momentum) standard deviation is $36\ (89)$ times the zero-point amplitude $q_{\rm zpf} \ (p_{\rm zpf})$.
This is, to the best of our knowledge, the first experimental  
demonstration of mechanical squeezing near quantum regimes using a macroscopic pendulum, whose mass-scale is high enough to measure gravitational interactions \cite{miliG1, gG1}. %% which will 

\section*{Theory for conditional squeezing}

\begin{figure}
\centering
\includegraphics[width=85mm]{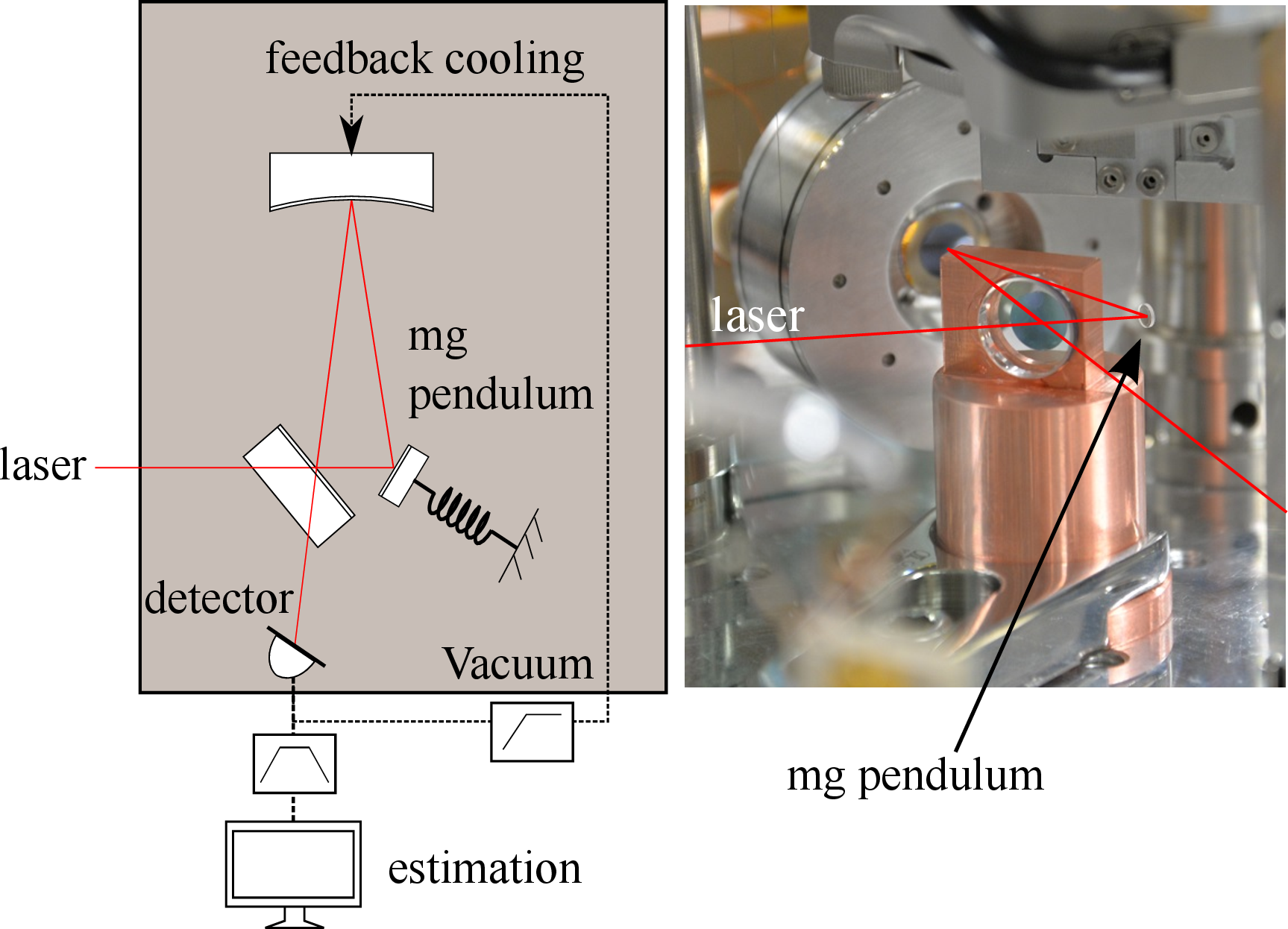}
\caption{(Color online) Experimental setup. 
The configuration of the three mirrors can stably trap the mirror's motion \cite{matsumoto2014}. 
}
\label{fig1}
\end{figure}

We consider a detuned cavity comprised of a pendulum of mass $m$ under feedback cooling, as shown in Fig.\ \ref{fig1}. 
Laser light enters the cavity and receives an intensity shift proportional to the mechanical position, which is read out via direct photo detection and fed back to the pendulum for cooling \cite{matsumoto2016}. 
We analyze the linearized Hamiltonian in a rotating frame at the laser frequency $\omega_L$, given by $H=\hbar\Omega(q^2+p^2)/4-\hbar\Delta(x^2+y^2)/4+\hbar gxq.$  
Here, $\hbar$ is the reduced Planck constant, $\Delta$ is the detuning of the optical cavity, $\Omega/2\pi$ is the bare mechanical resonance frequency, $x\ (y)$ is the dimensionless amplitude (phase) quadrature of the light, and $q\ (p)$ is the dimensionless position (momentum) of the mechanical oscillator.
$g\equiv G\sqrt{n_c}\sqrt{\hbar/2m\Omega}$ is the light-enhanced optomechanical coupling constant \cite{law}, where $G$ is the optical frequency shift per displacement, and $n_c$ is the number of photons circulating inside the cavity. 
The commutation relations are normalized as $[x,y]=[q,p]=2i$, resulting in the variance of each zero point motion to unity.

Under the adiabatic limit $(\kappa\gg\omega)$ and considering a small detuning $\Delta\ll\kappa$, we obtain the following quantum Langevin equations by adiabatically eliminating the cavity mode:
% \cite{sup}
\begin{eqnarray}
\dot{q}&=&\omega_m p, \nonumber\\
\dot{p}&=&-\omega_m q-\gamma_m p+\sqrt{2\gamma_m}p_{\rm in}-\frac{4g_m}{\sqrt{\kappa}}x_{\rm in}+\frac{8g_m\delta}{\sqrt{\kappa}}y_{\rm in}, \nonumber\\
X&=&-\frac{8g_m\delta\sqrt{\eta}}{\sqrt{\kappa}}q-\sqrt{\eta}x_{\rm in}+4\delta\sqrt{\eta}y_{\rm in}.
\label{eq2}
\end{eqnarray}
Here, $\kappa$ is the optical decay rate, $\omega_m$ is the mechanical resonance trapped in the optical potential, $\gamma_m$ is the mechanical decay rate under cooling, and $g_m\equiv g\sqrt{\Omega/\omega_m}=G\sqrt{n_c}x_{\rm zpf}$ is the coupling constant for the trapped mode. 
Further, $\delta\equiv\Delta/\kappa$ is the normalized detuning, $X$ is the measured optical amplitude quadrature with the detection efficiency $\eta$, and $x_{\rm in}$ and $y_{\rm in}$ ($p_{\rm in}$) refer to the optical (mechanical) noise input satisfying $\langle x^2_{\rm in}\rangle=\langle y^2_{\rm in}\rangle=2N_{\rm th}+1$ ($\langle p^2_{\rm in}\rangle=2n_{\rm th}+1$), where $N_{\rm th}$ ($n_{\rm th}$) is the thermal phonon number in light (the unconditional occupation under feedback cooling). 
From the third equation in Eq.~(\ref{eq2}), we can see that the measurement of light intensity can clearly provide linear continuous measurement of the position $q$, which induces the conditional mechanical squeezing of $q$; note that the quality of squeezing depends on the sensitivity coefficient defined by $A\equiv -8g_m\delta\sqrt{\eta}/\sqrt{\kappa}$ and the sensing (imprecision) noise components due to the optical noises. 
Note also that $q$ ($p$) is renormalized by a factor of $\sqrt{\Omega/\omega_m}$ ($\sqrt{\omega_m/\Omega}$) by taking into account the change in the resonance frequency by the optical spring.

To discuss how to evaluate the conditional squeezed state, we begin with describing the causal Kalman filter, which computes the least mean-square error estimate $\overrightarrow{q}(t)$ of the true value $q(t)$. 
$\overrightarrow{q}(t)$ is the expectation value conditioned on the measurement record (namely, $\overrightarrow{q}(t) \equiv E[q(t)|X(s), 0\leq s\leq t]$), which can be seen as {\it{prediction}} of the true value $q(t)$, based on the past data $\{X(s)|0\leq s\leq t\}$. 
The dynamics of the predicted value depends on the measurement record $X(t)$ and the conditional variance according to the Riccati equation \cite{ricatti, nurdin yamamoto}. 
When in a steady state, in the Fourier domain with the convention $F(\omega)=\int_{-\infty}^\infty f(t)\exp(i\omega t)dt$, the predicted value is calculated as $\overrightarrow{q}(\omega)=\overrightarrow{H}_q(\omega)X(\omega)$, where $\overrightarrow{H}_q(\omega)$ is the causal Wiener filter: 
\begin{eqnarray}
\overrightarrow{H}_{q}(\omega)=\frac{1}{\sqrt{\lambda_X M}}\frac{(\omega_X^2-\omega^2_m)-i\omega (\gamma_X-\gamma_m)}{F'(\omega)}. 
\label{eq3}
\end{eqnarray}
Here, $1/F'(\omega)=1/(\omega_X^2-\omega^2-i\omega\gamma_X)$ is a modified mechanical susceptibility, with resonance frequency $\omega_X=\sqrt[4]{\omega_m^4+2\Lambda_X \omega^3_m+\Gamma\lambda_X \omega_m^2}$ and decay rate $\gamma_X=\sqrt{\gamma^2_m-2\omega_m(\omega_m+\Lambda_{X})+2\omega^2_X}$. $\lambda_X$ is the measurement rate which determines the inverse time scale to resolve the zero-point fluctuation, $M=2\eta N_{th}+1$ is the total sensing noise, $\Lambda_X$ is related to the magnitude of back-action, and $\Gamma$ is the mechanical heating rate. 
The causal Wiener filter for the momentum, $\overrightarrow{H}_{p}(\omega)$, is given in Supplemental Material. 
The Wiener filter is the frequency-domain representation of the quantum Kalman filter based on the Langevin equation (\ref{eq2}), which maintains the Heisenberg uncertainty relation between $q$ and $p$, and $x$ and $y$.

The above state prediction procedure leads to quantum squeezing in the sense that the conditional variances satisfy $\langle (q(t)-\overrightarrow{q}(t))^2\rangle < 1 < \langle (p(t)-\overrightarrow{p}(t))^2\rangle$. 
However, 
%for experiments with macroscopic systems, 
in quantum mechanics, the true values $(q(t), p(t))$ can never be experimentally determined. 
%due to sensing noise. 
To circumvent 
%avoid 
this essential difficulty and to experimentally verify the prepared conditional squeezing, we 
%also 
calculate the conditional state 
%in an anti-causal manner 
using the anti-causal filter in addition to the above-described causal filter. 
%following Rossi {\it et al} \cite{verification}. 
This process, known as {\it retrodiction} \cite{verification}, computes the estimate $\overleftarrow{q}(t)$ for the true value $q(t)$ using the future data $\{X(s)|t\leq s\leq T\}$ after the entire measurement process is complete. 
Using the Kalman filter and the associated Riccati equation in an anti-causal manner, we can derive the frequency-domain retrodiction as 
$\overleftarrow{q}(\omega)=\overleftarrow{H}_q(\omega)X(\omega)$, where $\overleftarrow{H}_{q}(\omega)$ is the anti-causal Wiener filter:
\begin{eqnarray}
\overleftarrow{H}_{q}(\omega)=\frac{1}{\sqrt{\lambda_X M}}\frac{(\omega_X^2-\omega^2_m)+i\omega (\gamma_X+\gamma_m)}{F'(\omega)^*}.
\label{eq5}
\end{eqnarray}
Then, the conditional variance for the position, $V_{11}$, can be determined by comparing the results of prediction and retrodiction filters as follows; 

\begin{eqnarray}
\label{eq001}
\int^{\infty}_{-\infty}\frac{1}{2\pi}\big(|\overleftarrow{H}_{q}(\omega)|^2-|\overrightarrow{H}_{q}(\omega)|^2\big)S_{XX}(\omega)d\omega\approx 2 V_{11}. 
\end{eqnarray}
$S_{\overleftarrow{q}\overleftarrow{q}}(\omega)
=|\overleftarrow{H}_{q}(\omega)|^2 S_{XX}(\omega)$ and 
$S_{\overrightarrow{q}\overrightarrow{q}}(\omega)
=|\overrightarrow{H}_{q}(\omega)|^2 S_{XX}(\omega)$ 
are the power spectral density (PSD) of $\overleftarrow{q}(t)$ and $\overrightarrow{q}(t)$, respectively, where $S_{XX}(\omega)$ is the PSD of $X(t)$. 
Similarly, the conditional variance for the momentum, $V_{22}$, is calculated as 
\begin{eqnarray}
&\int^{\infty}_{-\infty}\frac{1}{2\pi}\big(|\overleftarrow{H}_{p}(\omega)|^2-|\overrightarrow{H}_{p}(\omega)|^2\big)S_{XX}(\omega)d\omega\approx 2 V_{22}.
\label{eq002}%\\
\end{eqnarray}
From these expressions, we can experimentally verify the conditional variances and accordingly the conditional squeezing, using the PSDs which can be constructed only from the output time series of $X(t)$. 
The details of the above discussion, including the validity of the approximation in Eqs.~(\ref{eq001}) and (\ref{eq002}), are given in Supplemental Material.

\section*{System characterization and position measurement}

The optical ring cavity with a cavity decay rate of $\kappa=2\pi\times1.64(2)$\ MHz consists of a 7.71(1) mg suspended mirror (measured by an accurate electronic balance: AND, BM-5), a 261.42(1) g suspended mirror, and a mirror fixed to a copper monolithic holder with its fundamental resonance at about 10 kHz. 
The bare mechanical dissipation for the mg pendulum ($\gamma_0$) and the heavier pendulum are $\gamma_0=2\pi\times4.74(5)\times10^{-5}$\ Hz and $2\pi\times 1.5\times10^{-4}$\ Hz, respectively. 
We can ignore the dynamics of the heavy and fixed mirrors because both have sufficiently small optomechanical coupling and Brownian motion amplitudes. 

The intrinsic dissipation of the mg pendulum shows a frequency dependence of $\gamma_0(\omega)\propto\gamma_0(\Omega)\times\Omega/\omega$, which is a characteristic referred to as structural damping \cite{saulson}. 
Thus the dissipation at the optical spring resonance frequency decreases as the resonance frequency increases. 
Note that due to the normal-mode splitting by the optical spring \cite{miliG1, sugiyama}, it can be reduced, in our case by a factor of 4. 
Without modification due to the optical spring and feedback cooling, the bare quality factor given by $\Omega/\gamma_0(\Omega)$ relates the resonance frequency to the magnitude of the thermal force noise ($\propto \gamma_0$). 
Thus, it can quantify the decoupling of the mechanical resonator from a thermal bath, leading to the number of coherence oscillation before the thermalization given by $\Omega/\gamma_0(\Omega) n_{\rm th}$ \cite{aspelmeyer}. 
To observe coherent oscillations more than unity, $Q\Omega>k_BT/\hbar$ should be satisfied, where $T$ is the room temperature. 
With the modification, the effective quality factor can be defined by $\omega_m/\gamma_0(\omega_m)$ or $\omega_m/\gamma_m$. 
In this case, the thermal noise is proportional to $\gamma_m n_{\rm th}$ as in (\ref{eq001}). 
The value of $\gamma_m n_{\rm th}$ is independent of the magnitude of feedback cooling because it determines the mechanical heating rate by a thermal bath, while the value of $n_{\rm th}$ can be reduced by feedback.  
Thus, the number of coherent oscillation is modified to be $\omega_m/\gamma_m n_{\rm th}$, leading to the condition for observing coherent oscillations given by $\omega_m\ \omega_m/\gamma_0(\omega_m)>k_B T/\hbar$. 
In \cite{miliG1} the value of the former quality factor given by $\omega_m/\gamma_0(\omega_m)$ reaches $10^8$, and moreover we show that the number of coherent oscillation can exceed unity within the state-of-the-art technology \cite{Q2}. 
On the other hand, the latter quality factor only represents the sharpness of the peak and it is independent of the thermal noise level.

Laser light (Coherent, Mephisto 500) is injected into the cavity with an incident laser power of 30 mW, and the reflected light is 
directly detected by a photo-detector  (HAMAMATSU, G10899-03K) of 92(2) \% efficiency. 
The efficiency is inferred by characterization of the optical spring. Its error includes both the error of the resistance in a current to voltage converter, and that of the incident laser power. 
To characterize the optomechanical interaction and the detection efficiency, we performed an auxiliary measurement to measure the resonance frequency of the optically trapped pendulum with varying detuning (cyan in Fig. 5 in Supplemental Material). 
The details of the auxiliary measurement can be found in Section~4 in Supplemental Material. 
We infer the frequency shift per displacement $G$ to be $-2\pi\times4.72(3)$\ PHz/m and extract the efficiency from the fitting of the measured resonance to the theory, given in Supplemental Material:
\begin{eqnarray}
\omega_m=\sqrt{\frac{8 \hbar G^2 n_c \delta} { (1+4\delta^2) \kappa m}}.
\label{os}
\end{eqnarray}

\begin{figure}[h]
%\hspace{-15mm}
\centering
\includegraphics[width=65mm]{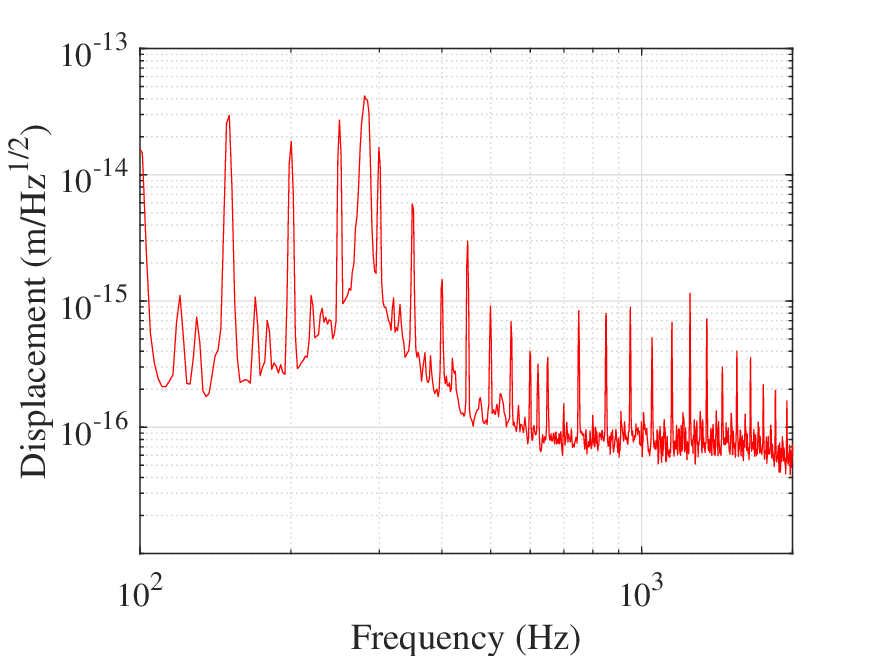}
\caption{(Color online) %The displacement spectrum with the mode temperature of 420 ${\rm \mu K}$ (blue) and 11 mK (red).
The displacement spectrum with the mode temperature of 11 mK (red). 
}
\label{fig10}
\end{figure}

Figure~\ref{fig10} shows the measured displacement spectrum with the unconditional mode temperature of the optically trapped pendulum of 11(2) mK, which is obtained using Welch’s method \cite{welch} with 1.9 Hz resolution and 50\% overlap. 
The phonon number $n_{\rm th}$ is correspondingly $8(2)\times10^5$, and the quality factor ($\equiv\omega_m/\gamma_m$) is $250(13)$. 
This data was calibrated to displacement (the meter-to-voltage conversion factor is $-2.3(4)\times10^{-10}\ {\rm m/V}$) based on a transfer function analysis, which can be independently derived from the value of the detuning (see Eq. 5 in \cite{matsumoto2016}). 

The cavity length was detuned from resonance such that the pendulum's resonance ($\omega_m/2\pi$) increases to 280(7) Hz. 
Because of the nonlinearity of the optical spring in Eq.~\ref{os} with respect to $\delta$, there are two possible values for the cavity detuning. 
We determined the value by analysing the data in the time domain (details can be seen in Section~5 in Supplemental Material. 
As a result, the mean detuning value is about $\Delta=0.03\times\kappa$, and the number of photons in the cavity is $1.16(7)\times10^{10}$, leading to a light-enhanced optomechanical coupling constant $g_m$ of $-2\pi\times3.2(2)\times10^4$\ Hz. 
Therefore, we obtain a quantum cooperativity 
($C_q \equiv C/n_{\rm th}$) of $0.0027(8)$. 
Furthermore, the sensitivity coefficient obtained is $A=14(1)\ \sqrt{\rm Hz}$.
This value is consistent with the meter-to-voltage conversion factor, which is measured independently from the sensitivity coefficient. 
We also consider contamination in the signal by laser classical noise \cite{miliG1};
The relative intensity noise level is $1.8\times10^{-8}\ {\rm /\sqrt{Hz}}$, which is 4.4 times higher than that of the shot noise limit. 
Thus, this contribution can be modeled as $N_{\rm th}=19$. 
The optomechanical parameters for the optimal state estimation are summarized in Table \ref{Table 1}.

\begin{table}[htb]
\begin{center}
    \caption{Optomechanical parameters}
    \label{Table 1}
    \begin{tabular}{|l|c|r||r|} \hline
      Parameter & Value \\ \hline \hline
      Mass &  $m=7.71(1)$\ mg \\
      %Mechanical (bare) dissipation& $\gamma=2\pi4.74(5)\times10^{-5}$\ Hz  \\
      Cavity decay rate & $\kappa=2\pi1.64(2)$\ MHz  \\%2pi822(10) Hz for amplitude
      Cavity detuning & $\Delta=0.03\times\kappa$ \\%0.0584(8) amplitude
      Circulating photon number& $n_c=1.17(6)\times10^{10}$ \\%5.44%
%      Frequency shift per displacement & $G=-2\pi4.72(3)$\ ${\rm PHz}/{\rm m}$ \\
      Light-enhanced coupling & $g_m=-2\pi3.2(2)\times10^4$\ Hz \\%5.83%
%    Multiphoton cooperativity & $C=2.2(2)\times10^{3}$ \\ %17.66%
%      Phonon occupancy & $n_{\rm th}=8(2)\times10^5$ \\ %3.97%
      Quantum cooperativity & $C_q=0.003$ \\ %30.86%
      Sensitivity coefficient & $A=14(1)\ \sqrt{\rm Hz}$ \\  %9%
      Excess noise in the laser & $N_{\rm th}=19$ \\ \hline %3.97%
\end{tabular}
\end{center}
\end{table}

\begin{figure}[h]
%\hspace{-15mm}
\centering
\includegraphics[width=70mm]{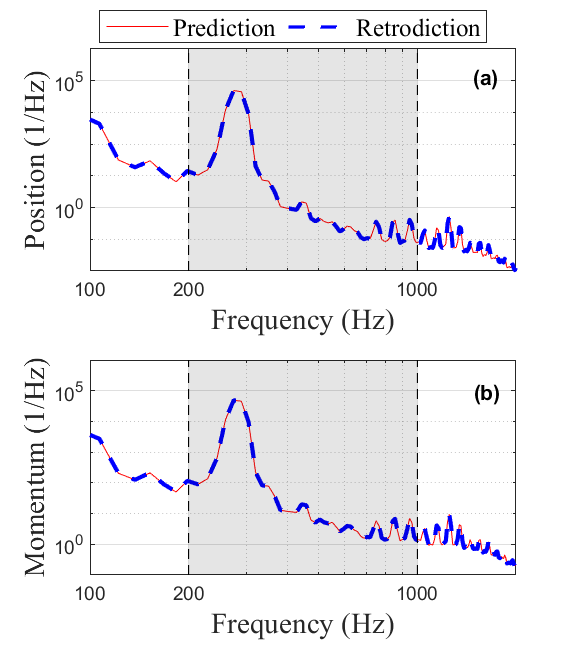}
\caption{(Color online)
PSD normalized by the zero-point amplitude (a) showing the prediction (red) and retrodiction (blue) for displacement. Plot (b) shows the prediction (red) and retrodiction (blue) for momentum. 
The shaded region represents the region where integration is calculated. 
}
\label{sfig3}
\end{figure}

\section*{Mechanical squeezing and discussions}
Here, we present mechanical squeezing using the data shown in Fig.~\ref{fig10}. 
Firstly, the $50$ Hz harmonics from the power supply are rejected by 1st order Butterworth notch filters with a 3 dB stop bandwidth of a few Hertz. 
Secondly, prediction and retrodiction are performed by multiplying the causal and anti-causal Wiener filters with the dimensionless (non-calibrated) amplitude quadrature $X$. 
The result is shown in Fig.~\ref{sfig3}, where the red lines show the PSD by prediction, $S_{\overrightarrow{q}\overrightarrow{q}}(\omega)$ and 
$S_{\overrightarrow{p}\overrightarrow{p}}(\omega)$; the blue dotted lines show the PSD by retrodiction, $S_{\overleftarrow{q}\overleftarrow{q}}(\omega)$ and 
$S_{\overleftarrow{p}\overleftarrow{p}}(\omega)$. 
Here, the parameters in the susceptibility of the filters are correspondingly given by $\omega_{X}=2\pi\times704$\ Hz and $\gamma_{X}=2\pi\times1073$\ Hz. The other terms are respectively $M=36$, $\Lambda_X=-2\pi\times578$ Hz, $\lambda_X=2\pi \times 0.87$ Hz, $\gamma_m=2\pi\times 1.1$ Hz and $\omega_m=2\pi\times 280$ Hz. 
Lastly, we use Eqs. (\ref{eq001}) and (\ref{eq002}) to calculate the verified conditional variances for position $V_{11}$ and momentum $V_{22}$ as 
\begin{flalign}
\label{V11 and V22}
V_{11}&=(1.30\pm0.32)\times 10^3,\\
V_{22}&=(8.02\pm1.13)\times 10^3.
%\operatorname{Covariance\hspace{0.1cm}Verification}&=(0.00^{+9.78}_{-7.09})\times 10^4
\end{flalign}
%V_{12}&=(0.59\pm4.25)\times 10^4
In terms of standard deviation, including the units, each noise level is $36\times q_{\rm zpf}$ and $89\times p_{\rm zpf}$. 
The achieved squeezing level is about 5 times closer to the zero point motion compared to the previous research \cite{multimode}. 
Here, the integration is calculated with the range over 200 Hz to 1000 Hz (shown as the shaded regions in Fig.~\ref{sfig3}) and the frequency resolution of 55 Hz. 
This frequency bandwidth corresponds to the region where the output equation of $X(t)$ in Eq.~(\ref{eq2}) is correctly modeled. 
The first reason of the discrepancy is that our model includes frequency independent PSD of the thermal noise, although the data fits the model given by the structural damping. 
The effect is especially apparent at frequencies lower than the resonance frequency. 
The second reason is that our model does not include multi-mode mechanical states such as the pitching mode and violin modes. 
To overcome the above issues, Meng {\it et al.} \cite{multimode} demonstrated to enhance the squeezing level using the multi-mode Wiener filter including the thermal noise model with structural damping. 
Furthermore, Shichijo {\it et al.} \cite{multimode2} recently report the derivation of the multi-mode Wiener filter considering the pendulum and rotational modes. 
Note that the theoretical conditional covariance between $q$ and $p$, i.e., $V_{12}$ given by Eq. (12) in Supplemental Material, is 2386, corresponding to the squeezing angle of about -18 degrees. 
%In case for a high enough measurement rate, the conditional variance at the squeezing angle with minimum value can be verified because the conditional covariance converges to zero (i.e. the squeezing angle becomes zero). 

\begin{figure} [h] 
%20210418\hspace{10mm}
\includegraphics[width=85mm]{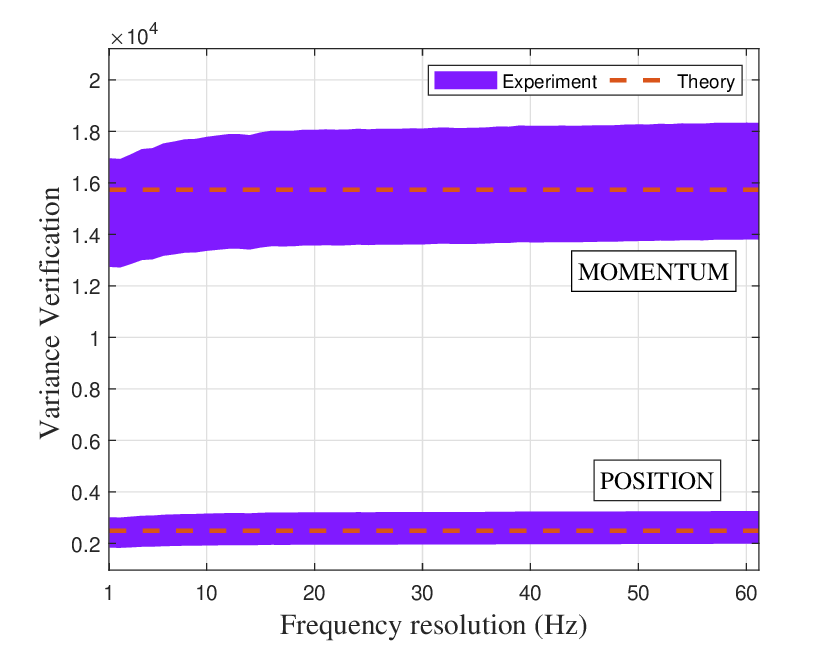}
\centering
\caption{Verification of the conditional mechanical squeezing. 
The orange dashed lines are the theoretical values of the conditional variances $2V_{11}$ and $2V_{22}$ for the position $q$ and momentum $p$, respectively. 
The purple shadows show the range of variances given by the modeling error.} 
\label{fig3}
\end{figure}

The 
%magnitude of the above 
errors in Eq.~(\ref{V11 and V22}) in the verification process come from the 
%is determined by 
modeling error. 
Since the optomechanical parameters are characterized within each specific uncertainties range (refer to Table \ref{Table 1}), these uncertainty 
%serve as potential candidates for generating 
may serve as the main source of modeling error. 
To numerically estimate the magnitude of the modeling errors, we conduct a Monte Carlo simulation; that is, we repeatedly generate the state with different parameters that are uniform randomly chosen according to Table \ref{Table 1}) in the causal and anti-causal filters with the frequency resolution from 1 Hz to 60 Hz. 
%By implementing a Monte Carlo simulation, we repeatedly verify the state with different parameters (randomly distributed according to Table \ref{Table 1}) in the causal and anti-causal filters to numerically estimate the magnitude of the modeling errors. 
%A 1-sigma error analyses is plotted in Fig. S4 in Supplemental Material \cite{sup}. %(For details, see Sumplemental Material \cite{sup}).\\
%Considering both aspects, we tuned the error in our optomechanical parameters to find the corresponding interval where the noise given by the model is inside the interval given by the simulation (see Table \ref{Table 2}). Then, we observe the corresponding interval in the frequency resolution where the verification is satisfied for position, momentum and covariance as shown in Fig. \ref{fig3}.
The results are presented in Fig.~\ref{fig3}. 
The experimentally verified conditional variances including the modeling errors, given in Eq.~(\ref{V11 and V22}), agree well with the theory. 
The tendency for discrepancy between the theory and experiment is observed at resolutions lower than 10 Hz. 
This is because that the resonance frequency of the pendulum fluctuates about 10 Hz due to fluctuations of the the cavity detuning (see Fig. 6 (c) in Supplemental Material. 
Since the resonance frequency $\omega_m$ is set as a constant in our model, the above fluctuations are not included in the model. 
Also, in the range beyond 60 Hz in the frequency resolution, the experimentally verified variances diverge, implying that the model is not anymore valid in this frequency regime. 
%which is caused by the difference between the model and experiment outside the integrated region. 
%The details on the error analysis are in section 6 in Supplemental Material \cite{sup}. %Given the good behaviour in the interval 10-60 Hz, we can focus in this interval to determine the validation of the verification process.

Our demonstration is the first step for generating entanglement between two massive pendulums in a PRFPMI via the radiation pressure of light \cite{entangle2, macroscopic}. 
To achieve this, the squeezed noise level must be less than the zero-point fluctuation, or in other words, the sensitivity of both the differential and common modes has to reach the SQL in a PRFPMI comprised of two pendulums. 
In general, interferometers can precisely measure the differential mode through common-mode noise rejection; however, the sensitivity for the common mode decreases significantly due to, e.g., laser frequency noise.
Since the result presented here is based on the direct measurement of the center-of-mass mode, it is possible to observe the common mode with the comparable sensitivity if two pendulums are combined in an interferometer. 
Moreover, we previously reported a monolithically constructed pendulum with lower dissipation \cite{Q2} that can satisfy the requirement for generating entanglement \cite{macroscopic}. 
%Figure~\ref{fig3}~(f) shows the theoretical prediction of the purity (convex upward) and　variance (noise level) of the squeezed quadrature (convex downward) of the conditional squeezed state, for the current experiment (red) and the future plan with the low-dissipative pendulum (black); the parameters used are given in Table~\ref{Table 1} for the red lines, while the black lines are obtained with the parameters given in \cite{Q2}, where particularly $C_q$ is improved to $C_q>1$. 
%The dash-dotted (black) line clearly shows genuine quantum squeezing around $\Delta/\kappa\simeq 0.3$, where the normalized variance in the squeezed quadrature is strictly less than 1. 
Thus, the combination of the low-dissipative oscillator and the mechanical squeezing reported here will result in the generation of an entangled state of mg-scale pendulums, which can be used to probe effects like quantum decoherence of macroscopic objects involving gravitational interactions. 
The result presented here is also the very first step towards entanglement via Newtonian gravity, in order to test the quantum nature of the gravitational interaction \cite{QG1,QG2,QG3,QG4,QG5}.

\section*{Conclusion}
We analytically derived the causal and anti-causal Wiener filters for a suspended mirror trapped in a detuned optical cavity. 
By applying these filters to the result of precise displacement measurement by direct photo-detection, we experimentally verified the conditional squeezing with its position variance achieving $1.3\times 10^3$ with an initial occupation of $8\times10^5$, in other words, with an initial mode temperature of 11 mK. 
Since our system can precisely measure displacement of the center-of-mass mode directly without depending on common-mode noise rejection, this research can pave the way to generate quantum entanglement between two massive pendulums in an interferometer, where both the differential and common modes have to be sensed with the SQL sensitivity. 
%The model validation was made by consideration the verification process, which was implemented by using prediction and retrodiction of the Kalman Filter. We have successfully derived the quantum filter for retrodiction and confirmed that the verified state aligns well with the theory.
%The model validation was made in several aspects, particularly from the fact 
%that the Wiener filter solution with precisely identified parameters based on the characterization of the optical spring effect, 
%effectively maximizes the purity of the mechanical squeezed state. 
%This further allows us to use the model and predict that genuine quantum squeezing of a macroscopic oscillator is within reach of the current state-of-the-art quantum technologies, combined with the techniques presented in this paper. 
In conclusion, the mechanical squeezing presented in this paper is the first step for quantum control of massive oscillators, especially in entanglement generation between massive pendulums, aiming towards the probing of unexplored phenomena such as gravity decoherence and the quantum nature of the gravitational interaction. 
\newpage

Supplemental Material
\setcounter{equation}{0}
\section{The model system}
The original system composed of the mechanical oscillator, the optical cavity, and the feedback control is governed by the following quantum Langevin equations: 
\begin{align*}
& \hspace*{0.em}
     \dot{x} = -\kappa x/2 - \Delta y + \sqrt{\kappa}x_{\rm in}, 
\nonumber \\ & \hspace*{0.em}
     \dot{y} = -\kappa y/2 + \Delta x + \sqrt{\kappa}y_{\rm in} - 2gq,
\nonumber \\ & \hspace*{0.em}
     \dot{q} = \Omega p, 
\nonumber \\ & \hspace*{0.em}
     \dot{p} = -\Omega q - \gamma_0 p + \sqrt{2\gamma_0}p_{\rm in} - 2gx - \int_{-\infty}^{t} ds g_{\rm FB}(t-s)X(s), 
\end{align*}
where $(q, p)$ are the position and momentum operators of the mechanical oscillator and 
$a=x+iy$ is the annihilation operator of the cavity mode. 
Also $\langle x_{\rm in}^2 \rangle=\langle y_{\rm in}^2 \rangle 
= 2N_{\rm th}+1$ and $\langle p_{\rm in}^2 \rangle=2k_BT/\hbar\Omega+1$, 
where $T$ is the room temperature. 
%In our experimental setup, the amplitude component of the output light field, $x_{\rm out}$, 
%is measured, leading to the following output equation: 
$g_{\rm FB}$ is a causal high-pass filter for cold damping (i.e. cooling by feedback) \cite{matsumoto2016}, which allows us to ignore the dissipative optical force applied on the mechanical oscillator (i.e., the imaginary part of the optical spring). 
Note that the laser frequency is not locked to the cavity via measurement-based (active) feedback but it is passively locked via the optical spring and the optical torsional spring effect.
$X$ is the measured output light field given by the following output equation: 
\begin{equation*}
     X= \sqrt{\eta}\, x_{\rm out} + \sqrt{1-\eta} \, x'_{\rm in},~~~
     x_{\rm out} = x_{\rm in} - \sqrt{\kappa}x, 
\end{equation*}
where $\eta \in [0,1]$ is the detection efficiency; 
note that $x'_{\rm in}$ is a fictitious vacuum field satisfying 
$\langle x'_{\rm in}\mbox{}^2 \rangle = 1$, introduced to represent 
imperfect detection. 
Now assume $\kappa \gg \Omega$, meaning that the cavity mode changes much faster than the mechanical oscillator mode. 
This allows us to adiabatically eliminate the cavity mode, and the resulting equation of motion 
of the mechanical oscillator is given by 
\begin{align*}
& \hspace*{0.em}
     \dot{q} = \Omega p,
\nonumber \\ & \hspace*{0.em}
     \dot{p} = -\Big( \Omega + \frac{16\Delta g^2}{\kappa^2 + 4\Delta^2}\Big) q 
                    - \gamma_m p 
                    + \sqrt{2\gamma_m}p_{\rm in} 
                     - \frac{4g \kappa \sqrt{\kappa} }{\kappa^2+4\Delta^2} x_{\rm in} 
                      + \frac{8g\Delta \sqrt{\kappa}}{\kappa^2+4\Delta^2} y_{\rm in},
\nonumber \\ & \hspace*{0.em}
     X = - \frac{8\Delta g\sqrt{\kappa \eta}}{\kappa^2+4\Delta^2} \, q 
              - \sqrt{\eta} \cdot \frac{\kappa^2-4\Delta^2}{\kappa^2+4\Delta^2} \, x_{\rm in} 
              + \frac{4\Delta \kappa \sqrt{\eta}}{\kappa^2+4\Delta^2} \, y_{\rm in}
              + \sqrt{1-\eta} \, x'_{\rm in},
\end{align*}
where $\gamma_m$ is the effective mechanical damping rate under feedback.  Correspondingly, the effective temperature of the center-of-mass of the mechanical oscillator is reduced to $T\gamma_0/\gamma_m$. 
Here we introduce the following change of variables and parameters: 
\begin{equation*}
      q = q' \sqrt{\frac{\Omega}{\omega_{m}}}, ~~~
      p = p' \sqrt{\frac{\omega_{m}}{\Omega}}, ~~~
%     p_{\rm in} = p_{\rm in}' \sqrt{\frac{\omega_{m}}{\Omega}}, ~~~
      \omega_m = \sqrt{\frac{16\Delta g^2}{\kappa^2+4\Delta^2}\Omega + \Omega^2}. 
\end{equation*}
The above system equations are then rewritten as 
\begin{align}
& \hspace*{0.em}
     \dot{q} = \omega_m p,
\nonumber \\ & \hspace*{0.em}
     \dot{p} = - \omega_m  q 
                    - \gamma_m p 
                    + \sqrt{2\gamma_m}p_{\rm in} 
                     - \frac{4g_m \kappa \sqrt{\kappa} }{\kappa^2+4\Delta^2} x_{\rm in} 
                      + \frac{8g_m\Delta \sqrt{\kappa}}{\kappa^2+4\Delta^2} y_{\rm in},
\nonumber \\ & \hspace*{0.em}
     X = - \frac{8\Delta g_m\sqrt{\kappa \eta}}{\kappa^2+4\Delta^2} \, q 
              - \sqrt{\eta} \cdot \frac{\kappa^2-4\Delta^2}{\kappa^2+4\Delta^2} \, x_{\rm in} 
              + \frac{4\Delta \kappa \sqrt{\eta}}{\kappa^2+4\Delta^2} \, y_{\rm in}
              + \sqrt{1-\eta} \, x'_{\rm in}, 
\label{Suppl Langevin}
\end{align}
where $g_m=g\sqrt{\Omega/\omega_m}$, and $(q', p')$ have been again represented by $(q, p)$ for simplifying 
the notation. 
This transformation of the mechanical resonance frequency also changes the autocorrelation of $p_{\rm in}$ to be $\langle p^2_{\rm in}\rangle=2n_{\rm th}+1$, where $n_{\rm th}=k_BT\gamma_0/\gamma_m \hbar\omega_m$ is the unconditional thermal occupation under feedback. 
Note that, unlike the system considered by Meng et al., in \cite{multimode}, both of the optical noise components $(x_{\rm in}, y_{\rm in})$ contribute to the dynamics and output equations.

%We now list the values of parameters used in our experimental system: 
%
\begin{comment}
\begin{align*}
& \hspace*{0.em}
      \omega_m = 2\pi \times 280~\mbox{Hz},~~~
      \gamma_m = 2\pi \times 1.1~\mbox{Hz},~~~
      \kappa = 2\pi \times 1.64 \times 10^6 ~\mbox{Hz},~~~
      \Delta = 0.0292 \kappa,~~~
      g_m = 2\pi \times 3.2 \times 10^4 ~\mbox{Hz},
\nonumber \\ & \hspace*{0.em}
      \eta = 0.92,~~~
       N_{\rm th}=19,~~~
       n_{\rm th}=8\times 10^5.
\end{align*}
%
\end{comment}

%\textbf{New information below}\\
%To consider the results produced in Meng et al., in \cite{Bowen2020}
To generalize the model, we also consider the measurement of the phase quadrature as in Miki et al. \cite{macroscopic}.
%As in \cite{miki2023generating}, we also consider the information in the phase quadrature Y.
The input-ouput relation for the phase quadrature is also expressed as
\begin{flalign}
Y=\sqrt{\eta}y_{out}+\sqrt{1-\eta}y_{in}', \hspace{0.5cm}y_{out}=y_{in}-\sqrt{\kappa}y.
\nonumber
\end{flalign}
Again, the equation above can be solved as follows:
\begin{flalign}
Y=\frac{4g_m\kappa\sqrt{\eta \kappa}}{\kappa^2+4\Delta^2}q-\sqrt{\eta}\frac{4\kappa\Delta}{\kappa^2+4\Delta^2}x_{in}-\sqrt{\eta}\frac{\kappa^2-4\Delta^2}{\kappa^2+4\Delta^2}y_{in}+\sqrt{1-\eta}y_{in}'
\end{flalign}
%Although we use the information in the amplitude quadrature X, we employ the phase quadrature Y to compare it with other previous results \cite{zhang2017prediction,meng2020mechanical}.
%%%%%%%%%%%%%%%%%%%%%%%%%%%%%%%%%%%%%%%%%%%%%%%%%%
%%%%%%%%%%%%%%%%%%%%%%%%%%%%%%%%%%%%%%%%%%%%%%%%%%
%%%%%%%%%%%%%%%%%%%%%%%%%%%%%%%%%%%%%%%%%%%%%%%%%%
Furthermore, we can rewrite the Langevin equation for both quadratures in matrix form: %\cite{macroscopic}:
\begin{flalign} 
\label{eq:uncond}%\label{langevin}
\frac{d\bm{r}}{dt}&=\bm{A r}+\begin{pmatrix}
0\\
w
\end{pmatrix},\nonumber\\
X&=\bm{C}_{X}\bm{r}+v_X,\\
Y&=\bm{C}_{Y}\bm{r}+v_Y,
\nonumber
\end{flalign}
where
\begin{flalign}
\bm{A}&=\begin{pmatrix}
0 &\omega_m\\
-\omega_m & -\gamma_m
\end{pmatrix},\hspace{0.5cm} w=\sqrt{2\gamma_m}p_{in}-\frac{4g_mk^{3/2}}{\kappa^2+4\Delta^2}x_{in}+\frac{8g_m\kappa^{1/2}\Delta}{\kappa^2+4\Delta^2}y_{in}\nonumber\\
\bm{C}_X&=\begin{pmatrix}
-\frac{8g_m\Delta\sqrt{\eta\kappa}}{\kappa^2+4\Delta^2} &0\nonumber\\
\end{pmatrix},\hspace{0.5cm} v_X=-\frac{\kappa^2-4\Delta^2}{\kappa^2+4\Delta^2}\sqrt{\eta}x_{in}+\frac{4\kappa \Delta}{\kappa^2+4\Delta^2}\sqrt{\eta}y_{in}+\sqrt{1-\eta}x_{in}'\nonumber\\
\bm{C}_Y&=\begin{pmatrix}\frac{4 g_m \kappa \sqrt{\eta \kappa}}{\kappa^2+4\Delta^2}& 0\end{pmatrix},\hspace{0.5cm} v_Y=-\frac{4\kappa \Delta}{\kappa^2+4\Delta^2}\sqrt{\eta}x_{in}-\frac{\kappa^2-4\Delta^2}{\kappa^2+4\Delta^2}\sqrt{\eta}y_{in}+\sqrt{1-\eta}y_{in}'\nonumber
\end{flalign}
Also, $\bm{r}=(q, p)^{T}$ is the vector for position and momentum of the mechanical system; $v_x, v_y$ describe the total sensing noise level given by the laser noise; $w$ represents the total force noise due to the Brownian motion and the back-action. 
The variances for the stochastic noises in the Langevin equations are computed in the following way:
\begin{flalign}
\langle w^2 \rangle &=2\gamma_m (2 n_{th}+1)+\frac{16 g_m^2 \kappa}{\kappa^2+4\Delta^2}(2 N_{th}+1)\equiv \Gamma,\nonumber\\
\langle w v_X \rangle &= \frac{4 g_m \kappa \sqrt{\kappa \eta}}{\kappa^2 +4 \Delta^2}(2 N_{th}+1),\nonumber\\
\langle w v_Y\rangle &=\frac{8g_m\Delta\sqrt{\kappa\eta}}{\kappa^2+4\Delta^2}(2N_{th}+1),\nonumber\\
\langle v_X^2\rangle &= \langle v_Y^2\rangle = (2\eta N_{th}+1). \nonumber
\end{flalign}
    %In Miki et al. \cite{macroscopic},  a quantum filter is employed to obtain the conditional state of the mechanical system. 
    Since the quantum system presented here only includes Gaussian noise, it is feasible to use the Kalman filter. %\cite{belavkin1995}. 
    The Kalman filter allows for the description of two type of outcomes: prediction and retrodiction. 
    The former is understood as the conditional estimation in a causal manner, while the latter refers the conditional estimation in an anti-causal manner. 
    %In this context, the Kalman Filter can provide information about the conditioned state using data from the measurement record $X$ or $Y$. \\
    %The vector $\bm{\overrightarrow{r}}=(\overrightarrow{q},\overrightarrow{p})^T$ represents the conditional state  
    %and
    %$\bm{V}=\langle (r-\overrightarrow{r})(r-\overrightarrow{r})^{T}\rangle$ is the covariance matrix that includes the information about the conditional variance for position and momentum.
 $\bm{\overrightarrow{r}}=(\overrightarrow{q},\overrightarrow{p})^T$  and $\bm{V}=\langle (r-\overrightarrow{r})(r-\overrightarrow{r})^{T}\rangle$ represent respectively the first and second moment for the predicted conditional state while $\bm{\overleftarrow{r}}=(\overleftarrow{q},\overleftarrow{p})^T$  and $\bm{V}_E=\langle (r-\overleftarrow{r})(r-\overleftarrow{r})^{T}\rangle$ denote the first and second moment for retrodictive conditional state.
 The differential equation for the conditional variance $\bm{V}$ corresponds to the following Riccati differential equation:
\begin{flalign}
\frac{d \bm{V}}{dt}=\bm{AV+V}\bm{A}^{T} + \bm{N}-(\bm{V}\bm{C}_{I}^T+\bm{L}_{I})M^{-1}(\bm{V}\bm{C}_{I}^T+\bm{L}_{I})^T,
\end{flalign}
where $I$ represents either $X$ or $Y$. 
Furthermore, we define:
\begin{flalign}
    \bm{L}_I&=\begin{pmatrix}
        0 \\ \langle wv_{I}\rangle
\end{pmatrix},\nonumber\\
    \bm{V} &=\begin{pmatrix}
        V_{11}&V_{12} \\
        V_{12}&V_{22}
\end{pmatrix},\hspace{0.5cm}
    \bm{N} =\begin{pmatrix}
        0 & 0\\
        0 & \langle w^2 \rangle
    \end{pmatrix}\nonumber\\
    M&=\langle v^2_I\rangle=2\eta N_{th}+1\nonumber
\end{flalign}
 The solution for the stationary case $\dot{\bm{V}}=0$ reads \cite{macroscopic}:
\begin{flalign}
V_{11}&=\frac{\gamma_I-\gamma_m}{\lambda_I}\nonumber\\
V_{12}&=\frac{V^2_{11}}{2\omega_m}\lambda_I\label{eq:cond}\\
V_{22}&=\frac{V_{11}}{2\omega^2_m}(2\omega_m(\omega_m+\Lambda_I)+\gamma_I V_{11}\lambda_I)\nonumber
\end{flalign}
The solution $V_{11}$ ($V_{22}$) represents the predicted conditional variance for position (momentum). $V_{12}$ is the predicted conditional covariance between position and momentum. %We use the term 'prediction' conditional variance to distinguish it from the 'retrodiction' conditional variance obtained below in the verification protocol. 
Moreover, $\gamma_I$ and $\omega_I$ are expressed as
\begin{flalign}
\gamma_I=\sqrt{\gamma^2_m-2\omega_m(\omega_m+\Lambda_{I})+2\omega^2_I},
\nonumber
\end{flalign}
\begin{flalign}
\omega_I=\sqrt[4]{\omega_m^4+2\Lambda_I \omega^3_m+\Gamma\lambda_I \omega_m^2}.
\nonumber
\end{flalign}
%The physical meaning of $\gamma_I$ and $\omega_I$ is related with the modified susceptibility produced by the measurement process. 
Here, we can define the measurement rate $\lambda_I=\bm{C}_{I}\bm{C}^{T}_{I}M^{-1}$, which represents the inverse time scale to spatially resolve the zero-point motion. 
Also, as $\Lambda_I=\sqrt{\bm{C}_I \bm{C}^T_I \bm{L}^T_I \bm{L}_I}M^{-1}$ increases, the conditional position variance decreases, meaning that the $q$-squeezed state is more enhanced. 
The respective expressions for $\lambda_I$ and $\Lambda_I$ are given by
\begin{flalign}
\lambda_X&=\frac{64g_m^2\eta\kappa\Delta^2}{(2\eta N_{th}+1)(\kappa^2+4\Delta^2)^2},\nonumber\\   \lambda_Y&=\frac{16g_m^2\eta\kappa^3}{(2\eta N_{th}+1)(\kappa^2+4\Delta^2)^2},\nonumber\\
    \Lambda_X&=-\Lambda_Y=-\frac{32g_m^2\eta\kappa^2\Delta}{(\kappa^2+4\Delta^2)^2}\frac{2N_{th}+1}{2\eta N_{th}+1}.
\nonumber
\end{flalign}
In the context of Meng et al. \cite{multimode} where $N_{th}=0$, $\Delta=0$ and $I=Y$, we have $\lambda_Y=4C\gamma_m \eta$. Here $C=4 g_m^2/(\kappa\gamma_m)$ is defined as cooperativity. 
%\textcolor{blue}{The value for $\lambda_Y$ corresponds to the calibration factor between the phase quadrature of the light and position quadrature of the mechanical motion.}
We also see that if $\lambda_I=0$ and $\Lambda_I=0$, then $\omega_I=\omega_m$ and $\gamma_I=\gamma_m$, which recovers the initial values of the mechanical susceptibility.
% like in the case established for $I=Y$, $\Delta=0$ and $N_{th}=0$ \cite{rossi2019observing}

%For the predictive state $\bm{\overrightarrow{r}}=(\overrightarrow{q},\overrightarrow{p})^T$, we have the following differential equation:

%\begin{flalign}
%\dot{\bm{\overrightarrow{r}}}=\bm{A\overrightarrow{r}+(VC^{T}_{I}+L_{I})M^{-1}(I - C_I\overrightarrow{r})}
%\end{flalign}

%"$\overrightarrow{}$" will indicate the predicted state based on the measurement and the positive time evolution of the estimation. On the contrary, "$\overleftarrow{}$" will indicate the retrodiction state based on the measurement and the negative time evolution of the estimation. By comparing prediction and retrodiction states, we will obtain the verification protocol.

%\begin{flalign}
%C_{0X}&=-\frac{8g_m \Delta \sqrt{\eta\kappa}}{\kappa^2+4\Delta^2},\hspace{0.5cm}
%C_{0Y}=\frac{4g_m \kappa \sqrt{\eta\kappa}}{\kappa^2+4\Delta^2}
%\end{flalign}
%We can see that $\lambda_I=C^{T}_{I}C_{I}M^{-1}$
%and $\Lambda_I=\sqrt{C^T_{I}C_I L_I L^T_I}M^{-1}$where $I= X$ or $Y$.\\

To further explore the implications of the modified susceptibility, we obtain the PSD for the output signal $I=X$ or $I=Y$. 
%Since we desired to explore the conditional state in frequency domain, 
We calculate the corresponding Fourier transformation $F(\omega)=\int_{-\infty}^\infty f(t)e^{i\omega t}dt$ for our Langevin equations in \eqref{eq:uncond}. 
Thus:
\begin{flalign}
-i\omega \bm{r}(\omega)=\bm{A} \bm{r}(\omega)+\begin{pmatrix}
    0\\w
\end{pmatrix}.
\nonumber
\end{flalign}
Then, we have
\begin{flalign}
q(\omega)=\frac{\omega_m}{F(\omega)}\bigg\{\sqrt{2\gamma_m}p_{in}(\omega)-\frac{4g_m \kappa \sqrt{\kappa}}{\kappa^2+4\Delta^2}x_{in}(\omega)+\frac{8\Delta g_m \sqrt{\kappa}}{\kappa^2+4\Delta^2}y_{in}(\omega)\bigg\}\nonumber
\end{flalign}
and
\begin{flalign}
    p(\omega) 
       = - \frac{i \omega}{\omega_m} q(\omega)
       = \frac{-i \omega}{F(\omega)} 
             \Big\{ \sqrt{2\gamma_m} p_{\rm in}(\omega)
                  - \frac{4g_m \kappa \sqrt{\kappa}}{\kappa^2+4\Delta^2} x_{\rm in}(\omega)
                  + \frac{8 \Delta g_m \sqrt{\kappa}}{\kappa^2+4\Delta^2} y_{\rm in}(\omega) \Big\}\nonumber,
\end{flalign}
where
\begin{flalign}
F(\omega)=\omega^2_m-i\gamma_m \omega-\omega^2.
\nonumber
\end{flalign}
The Fourier transformation for the linear measurement is expressed as 
\begin{flalign}
I(\omega)=\bm{C}_{I}\bm{r}(\omega)+v_I(\omega).
\nonumber
\end{flalign}
%Furthermore, we compute the Power Spectral Density. 
The symmetrized single-sided spectral density $S_{AB}(\omega)$ is defined through 
$2\pi \delta(\omega-\omega')S_{AB}(\omega)=\langle A(\omega)B^{\dagger}(\omega')+B^{\dagger}(\omega')A(\omega)\rangle$. 
For position and momentum, we obtain:
\begin{flalign}
S_{qq}(\omega)=\frac{\omega^2_m\Gamma}{|F(\omega)|^2}, \hspace{0.5cm}
S_{pp}(\omega)=\frac{\omega^2\Gamma}{|F(\omega)|^2}\nonumber
\end{flalign}
and their variances
\begin{flalign}
V_{qq}=\int^{\infty}_{-\infty}\frac{1}{2\pi} (S_{qq}(\omega))d\omega=\Gamma/2\gamma_m,\hspace{0.5cm}V_{pp}=\int^{\infty}_{-\infty}\frac{1}{2\pi} (S_{qq}(\omega))d\omega=\Gamma/2\gamma_m\label{psdq}.
\end{flalign}
%\begin{flalign}
%X(\omega)=\frac{-8g_m\Delta \sqrt{2\kappa\gamma_m\eta}}{(\kappa^2+4\Delta^2)F(\omega)}p_{in}(\omega)+\sqrt{\eta}\bigg\{\frac{32g_m^2\kappa^2\Delta\omega_m}{(\kappa^2+4\Delta^2)^2}-\frac{\kappa^2-4\Delta^2}{\kappa^2+4\Delta^2}\bigg\}x_{in}(\omega)+\sqrt{1-\eta}x^{'}_{in}(\omega)
%\end{flalign}
For $I= X, Y$, we calculate the PSD:
\begin{flalign}
S_{II}(\omega)&=C_I C_I^T S_{qq}(\omega)+S_{v_I v_I}(\omega)+\sqrt{C_I C_I^T}S_{q v_I}(\omega)+\sqrt{C_I C_I^T}S_{v_I q}(\omega)\nonumber
%S_{II}(\omega)&=C_IC^T_I S_{qq}+S_{v_I v_I}(\omega)+\frac{\omega_m \Lambda_{I}}{F(\omega)}+\frac{\omega_m \Lambda_{I}}{F^{*}(\omega)}\nonumber\\
%S_{II}(\omega)&=C_I C^T_I\frac{\omega_m \bar{n}}{|F(\omega)|^2 }+(2\eta N_{th}+1)\frac{|F(\omega)|^2}{|F(\omega)|^2}+2\frac{\omega_m \Lambda_I(2\eta N_{th}+1)(\omega^2-\omega_m^2)}{|F(\omega)|^2}\nonumber\\
%S_{II}(\omega)&=\frac{(2\eta N_{th}+1)}{|F(\omega)|^2}\big(\lambda_I \omega_m \bar{n} +((\omega^2-\omega_m^2)^2+\omega^2\gamma_m^2)^2+2\omega_m \Lambda_I(\omega^2-\omega_m^2)\big)
\end{flalign}
Then,
\begin{flalign}
S_{II}(\omega)=M\bigg(\frac{(-\omega^2+\omega_I^2)^2+\omega^2 \gamma_I^2}{(-\omega^2+\omega_m^2)^2+\omega^2\gamma_m^2}\bigg)\nonumber
\end{flalign}
or
\begin{flalign}\label{psdI}
S_{II}(\omega)=M\frac{|F'(\omega)|^2}{|F(\omega)|^2}
\end{flalign}
such that,
\begin{flalign}
    F'(\omega)=\omega^2_I-i\gamma_I \omega-\omega^2\nonumber
\end{flalign}
%The Power Spectral Density for the linear measurement shows us the division between two susceptibilities. 
Again, when no measurement is performed, with $\lambda_I=0$ and $\Lambda_I=0$, then $F'(\omega)=F(\omega)$ and the PSD transforms into white noise. In this scenario, the physical significance of the matrix $\bm{V}$ is lost.
%Additionally, we can express $S_{II}(\omega)$ as
%\begin{flalign}
%2\pi \delta (\omega-\omega')S_{II}(\omega)=\langle I(\omega)I^{\dagger}(\omega')+I^{\dagger}(\omega ')I(\omega)\rangle
%\end{flalign}
%then, we obtain
%\begin{flalign}
%I(\omega)=\big(\sqrt{2\eta N_{th}+1}\big)\bigg(\frac{\omega^2-\omega^2_I-i\omega \gamma_I}{\omega^2-\omega^2_m-i\omega\gamma_m}\bigg)\epsilon_I(\omega)
%\end{flalign}
%such that,
%\begin{flalign}
%\langle \epsilon_I(\omega)\epsilon^{\dagger}_I(\omega ') \rangle =2\pi\delta (\omega-\omega')
%\end{flalign}

\section{Quantum estimation in frequency domain}

In our analysis, we focus on the stationary behavior of the mechanical system. 
%Performing a non-stationary analysis would require additional considerations, such as accounting for other sources of noise and incorporating mathematical models for changes in the mechanical resonance frequency over time. 
Therefore, we can obtain a quantum causal Wiener filter by examining the quantum causal Kalman filter in the frequency domain. 
We transform $F(\omega)=\int_{-\infty}^\infty f(t)e^{i\omega t}dt$. 
Then, from the differential equation of the first moment $\overrightarrow{\bm{r}}$ \cite{ricatti}: 
\begin{flalign}
\dot{\overrightarrow{\bm{r}}}=\bm{A}\overrightarrow{\bm{r}}+(\bm{V}\bm{C}^{T}_{I}+\bm{L}_{I})M^{-1}(I - \bm{C}_I\overrightarrow{\bm{r}})
\end{flalign}
and using $\dot{\overrightarrow{\textbf{r}}}(t)=-i\omega\overrightarrow{\textbf{r}}(\omega)$, we simply obtain:
\begin{flalign}
-i\omega\overrightarrow{\bm{r}}(\omega)=\bm{A}\overrightarrow{\bm{r}}(\omega)+(\bm{V}\bm{C}^{T}_{I}+\bm{L}_{I})M^{-1}(I(\omega) - \bm{C}_I\overrightarrow{\bm{r}}(\omega)).
\nonumber
\end{flalign}
%and its respective Power Spectral Density:
%\begin{flalign}
%S_{II}(\omega)=(2\eta N_{th}+1)\frac{((\omega^2-\omega_I^2)^2+\omega^2\gamma_I^2)}{((\omega^2-\omega_m^2)^2+\omega^2\gamma_m^2)}
%\end{flalign}
%\begin{flalign}
%\overrightarrow{q}(\omega)=\frac{1}{C_{I}}\frac{(-\omega_m(V_{12}\lambda_{I}+\Lambda_{I})-V_{11}\lambda_I\gamma_m-i\omega V_{11}\lambda_I)I(\omega)}{(-\omega^2-i\omega(\gamma_m-V_{11}\lambda_{I})+\omega^2_m+(\omega_m(V_{12}\lambda_{I}+\Lambda_{I})+V_{11}\lambda_{I}\gamma_m))}
%\end{flalign}
We derive the steady-state solution for position and momentum in the frequency domain:
\begin{flalign}
\overrightarrow{q}(\omega)&=\frac{1}{\sqrt{\lambda_I M}}\frac{(\omega^2_I-\omega^2_m-i\omega V_{11}\lambda_I)I(\omega)}{F'(\omega)},
\nonumber\\
\overrightarrow{p}(\omega)&=\frac{1}{\sqrt{\lambda_I M}\omega_m}\frac{(-V_{11}\lambda_I \omega^2_m-i\omega(\omega^2_I-\omega^2_m+V_{11}^2\lambda^2_I-V_{11}\lambda_I\gamma_I))I(\omega)}{F'(\omega)}.
\label{prediction}
\end{flalign}
This is exactly the quantum causal Wiener filter that estimates the position $q$ and the momentum $p$ from the measurement record $I$ in the frequency domain. 
That is, we have:
\begin{flalign}
\overrightarrow{q}(\omega)&=\overrightarrow{H}_q(\omega)I(\omega), 
\nonumber\\
\overrightarrow{p}(\omega)&=\overrightarrow{H}_p(\omega)I(\omega),
\nonumber
\end{flalign}
%\end{flalign}
%then, we straightforwardly obtain:
%\begin{comment}
%\begin{flalign}
%\overrightarrow{H}_{q}(\omega)&=\frac{1}{\sqrt{\lambda_I M}}\frac{(\omega_I^2-\omega^2_m)-i\omega V_{11}\lambda_I}{F'(\omega)}\tag{S10}\\
%\overrightarrow{H}_{p}(\omega)&=\frac{1}{\sqrt{\lambda_I M}\omega_m}\frac{(-V_{11}\lambda_I \omega^2_m-i\omega(\omega^2_I-\omega^2_m+V_{11}^2\lambda^2_I-V_{11}\lambda_I\gamma_I))}{F'(\omega)}\tag{S11}
%\end{flalign}
%\end{comment}
%\begin{flalign}
where
\begin{flalign}
\overrightarrow{H}_{q}(\omega)&=\frac{1}{\sqrt{\lambda_I M}}\frac{(\omega_I^2-\omega^2_m)-i\omega (\gamma_I-\gamma_m)}{F'(\omega)},\\
\overrightarrow{H}_{p}(\omega)&=\frac{1}{\sqrt{\lambda_I M}\omega_m}\frac{(-(\gamma_I-\gamma_m) \omega^2_m-i\omega(\omega^2_I-\omega^2_m+(\gamma_I-\gamma_m)^2-(\gamma_I-\gamma_m)\gamma_I))}{F'(\omega)}.
\end{flalign}
%\overrightarrow{H}_{qp}(\omega)&=\overrightarrow{H}_{q}(\omega)\overrightarrow{H}^{*}_{p}(\omega)
For $\Delta=0$, $N_{th}=0$ and $I=Y$, $\overrightarrow{H}_{q}(\omega)$ and $\overrightarrow{H}_{p}(\omega)$ are equivalent to the filter obtained by Meng et al \cite{multimode}.
We can see that the following relations hold:
\begin{flalign}\label{eq:pcv}
V_{11}&=\int^{\infty}_{-\infty}\frac{1}{2\pi} (S_{qq}(\omega)-S_{\overrightarrow{q}\overrightarrow{q}}(\omega))d\omega\nonumber=V_{qq}-V_{\overrightarrow{q}\overrightarrow{q}},\\
%V_{11}&=\frac{\omega_m^2 \bar{n}}{\gamma_m \omega_m^2}-\frac{(\omega_I^2-\omega_m^2)^2+\omega_m^2\lambda_I^2 V^2_{11}}{\lambda_I \gamma_m \omega_m^2}\nonumber
V_{22}&=\int^{\infty}_{-\infty}\frac{1}{2\pi} (S_{pp}(\omega)-S_{\overrightarrow{p}\overrightarrow{p}}(\omega))d\omega=V_{pp}-V_{\overrightarrow{p}\overrightarrow{p}},\\
V_{12}&=\int^{\infty}_{-\infty}\frac{1}{2\pi}\operatorname{Re}\big(S_{qp}(\omega)-S_{\overrightarrow{q}\overrightarrow{p}}(\omega)\big)d\omega,
\nonumber
\end{flalign}
where we used the following equations:
\begin{equation*}
     \int_{-\infty}^\infty \frac{1}{|F(\omega)|^2}d\omega 
     = \frac{\pi}{\gamma_m \omega_m^2}, ~~
     \int_{-\infty}^\infty \frac{\omega^2}{|F(\omega)|^2}d\omega 
     = \frac{\pi}{\gamma_m}.
\end{equation*}
%\cite{weiss2005course}.
\section{Verification protocol}

In order to experimentally verify the conditional variance $\bm{V}$, we use the retrodiction process in addition to the prediction process, using the quantum anti-causal Kalman filter. 
Since the unconditional variances of the true variables of $q$ and $p$ are experimentally unavailable, we perform a retrodiction-based verification technique \cite{verification} in the frequency domain to experimentally access the value of the conditional variance. 
For a Gaussian system, it is also possible to use the quantum Kalman filtering as well as the previous case for prediction \cite{ricatti}:
%\begin{flalign}
%\dot{\overrightarrow{\bm{r}}}=\bm{A\overrightarrow{r}+(VC^{T}_{I}+L_{I})M^{-1}(I - C\overrightarrow{r})}
%\end{flalign}
%we have the following differential equation for retrodiction :
\begin{flalign}
\dot{\overleftarrow{\bm{r}}}
  =-\bm{A}\overleftarrow{\bm{r}}
     +(\bm{V}_{E} \bm{C}^{T}_{I} - \bm{L}_{I})M^{-1}(I - \bm{C}_I\overleftarrow{\bm{r}})
\end{flalign}

As defined in the previous section, $\bm{\overleftarrow{r}}=(\overleftarrow{q},\overleftarrow{p})^T$ represents the retrodictive conditional state  
    and
    $\bm{V_{E}}=\langle (r-\overleftarrow{r})(r-\overleftarrow{r})^{T}\rangle$ is the retrodictive covariance matrix.
The respective Riccati equation for retrodiction is expressed as given by Zhang et al. \cite{ricatti}
\begin{flalign}
\frac{d \bm{V}_{E}}{dt}=-\bm{A}\bm{V}_{E}-\bm{V}_{E}\bm{A}^{T}\bm{+N}-(\bm{V}_{E}\bm{C}_{I}^T-\bm{L}_{I})M^{-1}(\bm{V}_{E}\bm{C}_{I}^T-\bm{L}_{I})^T
\end{flalign}
%We solve the stationary case by setting $\bm{\dot{V}_{E}}=0$\\
%\begin{flalign}
%\frac{d \textbf{V}_{E}}{dt}=-\textbf{A}%\textbf{V}_{E}-\textbf{A}^{T}\textbf{V}_{E}%\textbf{+N}-(\textbf{V}_{E}\textbf{C}_{I}%^T-\textbf{L}_{I})^T\textbf{M}^{-1}%(\textbf{V}_{E}\textbf{C}_{I}^T-\textbf{L}%_{I})^T
%\end{flalign}
where $ I=X$ or $Y$, and $M=\langle v^2_X \rangle =\langle v^2_{Y}\rangle=(2\eta N_{th}+1)$. %Each matrix is given by
Here, ${V}_{E}$ is given by
\begin{flalign}
\bm{V}_{E}=\begin{pmatrix}
V_{E11} & V_{E12}\\
V_{E12} & V_{E22}
\end{pmatrix}.\nonumber \hspace{0.1cm} %\textbf{N}=\begin{pmatrix}
%0 & 0\\
%0 & \langle w^2 \rangle
%\end{pmatrix}\nonumber
\end{flalign}

Considering the steady state $\dot{\textbf{V}}_E=0$, the covariance matrix satisfies the following equations:\\
\begin{flalign}
-2\omega_m V_{E12}-V^{2}_{E11}\lambda_{I}=0\nonumber\\
\omega_m( V_{E11}-V_{E22})+ (\gamma_m -V_{E11}\lambda_{I})V_{E12}+V_{E11}\Lambda_{I}=0\nonumber\\
2\gamma_m V_{E22}+2\omega_m V_{E12} -(\sqrt{\lambda_I}V_{E12}-\Lambda_{I}/\sqrt{\lambda_I})^2+\Gamma=0\nonumber
\end{flalign}
After solving,
\begin{flalign}
V_{E11}&=\frac{\gamma_I+\gamma_m}{\lambda_I}\nonumber\\
V_{E12}&=-\frac{V^2_{E11}}{2\omega_m}\lambda_I\\
V_{E22}&=\frac{V_{E11}}{2\omega_m^2}(2\omega_m(\omega_m+\Lambda_I)+\gamma_IV_{E11}\lambda_I)\nonumber
\label{eq:retro}
\end{flalign}

The solution for the retrodictive conditional variance is similar to what was obtained before for prediction. The change in the direction of time for the estimation is reflected by the sign in front of the mechanical dissipation $\gamma_m$.
As with the previous approach, we solved the differential equation for $\bm{r}$ by introducing the Fourier transformation, $\bm{\dot{\overleftarrow{r}}} = i\omega\overleftarrow{r}$. However, it is important to note that the sign in the transformation is different due to the estimation going in the negative direction of time.
\begin{flalign}
\dot{\overleftarrow{\bm{r}}}=-\bm{A}\overleftarrow{\bm{r}}+(\bm{V}_{E}\bm{C}^{T}_{I}-\bm{L}_{I})M^{-1}(I - \bm{C}_I\overleftarrow{\bm{r}})
\end{flalign}
We straightforwardly obtain
\begin{flalign}
i\omega \overleftarrow{\bm{r}}(\omega)=-\bm{A}\overleftarrow{\bm{r}}(\omega)+(\bm{V}_{E}\bm{C}_{I}^T-\bm{L}_I)M^{-1}(I^{\dagger}(\omega)-\bm{C}_I\overleftarrow{\bm{r}}(\omega))
\nonumber
\end{flalign}
Then, calculating the value for position and momentum, we derive the following:
\begin{flalign}
\overleftarrow{q}(\omega)&=\frac{1}{\sqrt{\lambda_I M}}\frac{(\omega^2_I-\omega^2_m+i\omega V_{E11}\lambda_I)I^{\dagger}(\omega)}{F'(\omega)^{*}}\nonumber\nonumber\\
\overleftarrow{p}(\omega)&=\frac{1}{\sqrt{\lambda_I M}\omega_m}\frac{(V_{E11}\lambda_I \omega^2_m-i\omega(\omega^2_I-\omega^2_m+V_{E11}^2\lambda^2_I-V_{E11}\lambda_I\gamma_I))I^{\dagger}(\omega)}{F'(\omega)^{*}}\nonumber\nonumber
\end{flalign}
Again, we consider the mathematical representation of a filter applied over the measurement record.
\begin{flalign}
\overleftarrow{q}(\omega)=\overleftarrow{H}_{q}(\omega)I^{\dagger}(\omega)\nonumber
\end{flalign}
\begin{flalign}
\overleftarrow{p}(\omega)=\overleftarrow{H}_{p}(\omega)I^{\dagger}(\omega)\nonumber
\end{flalign}
then,
\begin{comment}
\begin{flalign}
\overleftarrow{H}_{q}(\omega)&=\frac{1}{\sqrt{\lambda_I M}}\frac{(\omega_I^2-\omega^2_m)+i\omega V_{E11}\lambda_I}{F'(\omega)^{*}}\tag{S17}\\
\overleftarrow{H}_{p}(\omega)&=\frac{1}{\sqrt{\lambda_I M}\omega_m}\frac{(V_{E11}\lambda_I \omega^2_m-i\omega(\omega^2_I-\omega^2_m+V_{E11}^2\lambda^2_I-V_{E11}\lambda_I\gamma_I))}{F'(\omega)^{*}}\tag{S18}
\end{flalign}
\end{comment}
\begin{flalign}
\overleftarrow{H}_{q}(\omega)&=\frac{1}{\sqrt{\lambda_I M}}\frac{ (\omega_I^2-\omega^2_m)+i\omega (\gamma_I+\gamma_m)}{F'(\omega)^{*}}\\
\overleftarrow{H}_{p}(\omega)&=\frac{1}{\sqrt{\lambda_I M}\omega_m}\frac{((\gamma_I+\gamma_m) \omega^2_m-i\omega(\omega^2_I-\omega^2_m+(\gamma_I+\gamma_m)^2-(\gamma_I+\gamma_m)\gamma_I))}{F'(\omega)^{*}}
\end{flalign}
In this case, we observe that the relationship between retrodictive conditional variance and unconditional variance differs from what was observed previously.
\begin{flalign}\label{eq:rcv}
V_{E11}&=\int^{\infty}_{-\infty}\frac{1}{2\pi} (S_{\overleftarrow{q}\overleftarrow{q}}(\omega)-S_{qq}(\omega))d\omega=V_{\overleftarrow{q}\overleftarrow{q}}-V_{qq}\nonumber\\
%&=\frac{(\omega_I^2-\omega_m^2)^2+\omega_m^2\lambda_I^2 V^2_{E11}}{\lambda_I \gamma_m \omega_m^2}-\frac{\omega_m^2 \bar{n}}{\gamma_m \omega_m^2}
V_{E22}&=\int^{\infty}_{-\infty}\frac{1}{2\pi} (S_{\overleftarrow{p}\overleftarrow{p}}(\omega)-S_{pp}(\omega))d\omega=V_{\overleftarrow{p}\overleftarrow{p}}-V_{pp}\\
V_{E12}&=\int^{\infty}_{-\infty}\frac{1}{2\pi}\operatorname{Re}\big(S_{\overleftarrow{q}\overleftarrow{p}}(\omega)-S_{qp}(\omega)\big)d\omega
=V_{\overleftarrow{q}\overleftarrow{p}}-V_{qp}. 
\nonumber
\end{flalign}
%We can see in this case that the law of total variance is given by, 
%\begin{flalign}
    %\bm{V}&=\int^{\infty}_{-\infty}e^{-\bm{A}\tau}\bm{S}\frac{\gamma_m \omega_m^2}{\pi}e^{-\bm{A}^{T}\tau}d\tau\\
    %\bm{V}&=-\bm{S}
%\end{flalign}

By comparing the equation above \eqref{eq:rcv} with the one obtained for predicted conditional variance \eqref{eq:pcv}, we have enough conditions for verification. The verification of the conditional variance can be derived from the following relations (\eqref{eq:rcv} + \eqref{eq:pcv})
\begin{flalign}
&\int^{\infty}_{-\infty}\frac{1}{2\pi}\big(S_{\overleftarrow{q}\overleftarrow{q}}(\omega)d\omega-S_{\overrightarrow{q}\overrightarrow{q}}(\omega)d\omega\big)=V_{11}+V_{E11}\nonumber\\
&\int^{\infty}_{-\infty}\frac{1}{2\pi}\big(S_{\overleftarrow{p}\overleftarrow{p}}(\omega)d\omega-S_{\overrightarrow{p}\overrightarrow{p}}(\omega)d\omega)\big)=V_{22}+V_{E22}\\
&\int^{\infty}_{-\infty}\frac{1}{2\pi}\operatorname{Re}\big(S_{\overleftarrow{q}\overleftarrow{p}}(\omega)d\omega-S_{\overrightarrow{q}\overrightarrow{p}}(\omega)d\omega\big)=V_{E12}+V_{12}\nonumber
\end{flalign}
When the measurement rate $\lambda_I$ is larger than $\gamma_m$, we can affirm the relations as Rossi et. al.\cite{verification}:
\begin{flalign}
    V_{E11}=\frac{\gamma_I+\gamma_m}{\lambda_I}=V_{11}+2\frac{\gamma_m}{\lambda_I}\approx V_{11}
    \label{eq:equi}
\end{flalign}
Consequently,
\begin{flalign}
    V_{E12}&=-\frac{V^{2}_{E11}\lambda_I}{2\omega_m}\approx -\frac{V^2_{11}}{2\omega_m}\lambda_I= -V_{12}\nonumber\\
    V_{E22}&=\frac{V_{E11}}{2\omega_m^2}(2\omega_m(\omega_m+\Lambda_I)+\gamma_I V_{E11}\lambda_I)\approx \frac{V_{11}}{2\omega^2_m}(2\omega_m(\omega_m+\Lambda_I)+\gamma_I V_{11}\lambda_I)=V_{22}\nonumber
\end{flalign}
Then,
\begin{flalign}
&\int^{\infty}_{-\infty}\frac{1}{2\pi}\big(S_{\overleftarrow{q}\overleftarrow{q}}(\omega)d\omega-S_{\overrightarrow{q}\overrightarrow{q}}(\omega) d\omega)\big)\approx 2 V_{11}\nonumber\\
&\int^{\infty}_{-\infty}\frac{1}{2\pi}\big(S_{\overleftarrow{p}\overleftarrow{p}}(\omega)d\omega-S_{\overrightarrow{p}\overrightarrow{p}}(\omega) d\omega)\big)\approx 2 V_{22}\\
&\int^{\infty}_{-\infty}\frac{1}{2\pi}\big(S_{\overleftarrow{q}\overleftarrow{p}}(\omega)d\omega-S_{\overrightarrow{q}\overrightarrow{p}}(\omega)d\omega )\big)\approx 0\nonumber
\end{flalign}
To find the conditional variances in each case, we can use predictive and retrodictive filters along with the PSD of the measurement record ($I=X$ for our case):
\begin{flalign}\label{eq:veri}
&\int^{\infty}_{-\infty}\frac{1}{2\pi}\big(|\overleftarrow{H}_{q}(\omega)|^2-|\overrightarrow{H}_{q}(\omega)|^2\big)S_{II}(\omega)d\omega\approx 2 V_{11}\nonumber\\
&\int^{\infty}_{-\infty}\frac{1}{2\pi}\big(|\overleftarrow{H}_{p}(\omega)|^2-|\overrightarrow{H}_{p}(\omega)|^2\big)S_{II}(\omega)d\omega\approx 2 V_{22}\\
&\int^{\infty}_{-\infty}\frac{1}{2\pi}\operatorname{Re}\big(\overleftarrow{H}_{q}(\omega)\overleftarrow{H}^*_{p}(\omega)-\overrightarrow{H}_{q}(\omega)\overrightarrow{H}^{*}_{p}(\omega)\big)S_{II}(\omega)\approx 0\nonumber
\end{flalign}

To ensure consistency with the previous report \cite{multimode}, we consider the following situation. 
When the measurement rate is larger than mechanical dissipation ($\lambda_I\gg \gamma_m$), the filters for prediction and retrodiction can be approximated as 
\begin{align}
    \overrightarrow{H}_q(\omega)\approx\overleftarrow{H}_q^{*}(\omega)\nonumber\\
    \overrightarrow{H}_p(\omega)\approx- \overleftarrow{H}^{*}_p(\omega)\nonumber. 
\end{align}
These are the same conditions as obtained in the previous research \cite{multimode}. 
Here, the time symmetry $t\rightarrow -t$ is restored in the quantum filters such that $\overrightarrow{q}(\omega)=\overleftarrow{q}^{*}(\omega)$ and $\overrightarrow{p}(\omega)=-\overrightarrow{p}^{*}(\omega)$. The minus sign in the momentum is a result of the time variation.

\newpage
\section{The auxiliary measurement}

\begin{figure}[h]
%\hspace{-15mm}
\centering
\includegraphics[width=70mm]{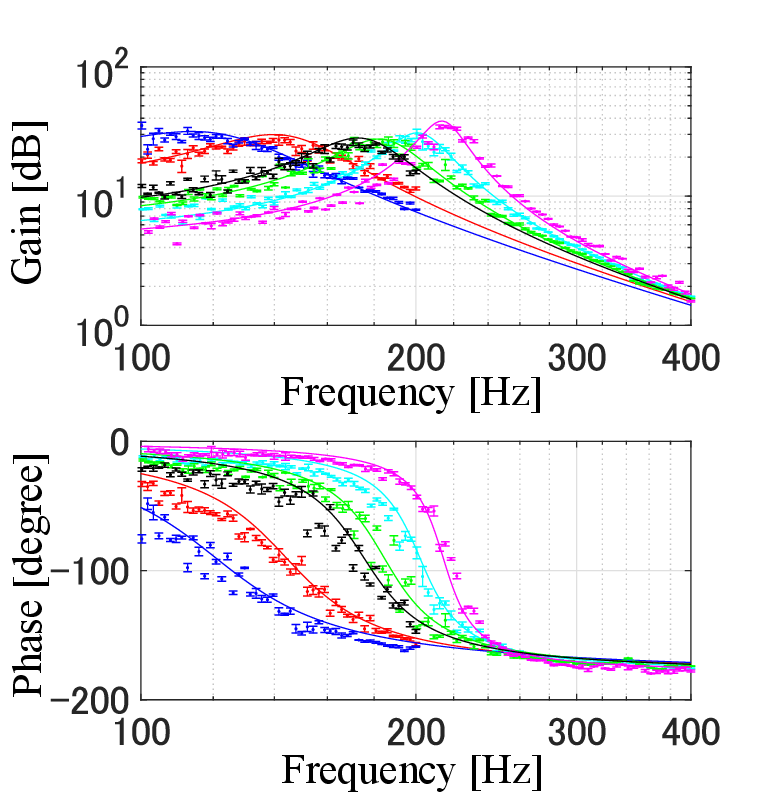}
\caption{(Color online) 
The gain and phase plots of the transfer function of the optically trapped mechanical pendulum. 
}
\label{sfig1}
\end{figure}

In order to characterize the optomechanical interaction, we measured the resonance frequency of the optically trapped pendulum with varying detuning. 
The resonance frequency of the optically trapped pendulum is given by
\begin{align}
    \omega_m=\sqrt{\frac{8\hbar G^2 n_c \delta}{(1+4\delta^2)\kappa m}}. 
\label{S8}
\end{align}
The resonance $\omega_m$ was identified by measuring the transfer function as shown in Fig.~\ref{sfig1}. 
This auxiliary measurement was performed with a relatively small incident laser power of 3 mW, compared to the main measurement of 30 mW in the main text. 
Thus, the results are compensated for the power difference by multiplying the measured resonance by the square root of the power ratio shown as cyan dots in Fig. \ref{sfig2} (a).

\section{Main data in the time domain}
\begin{figure} [h]
%\hspace{-10mm}
\includegraphics[width=80mm]{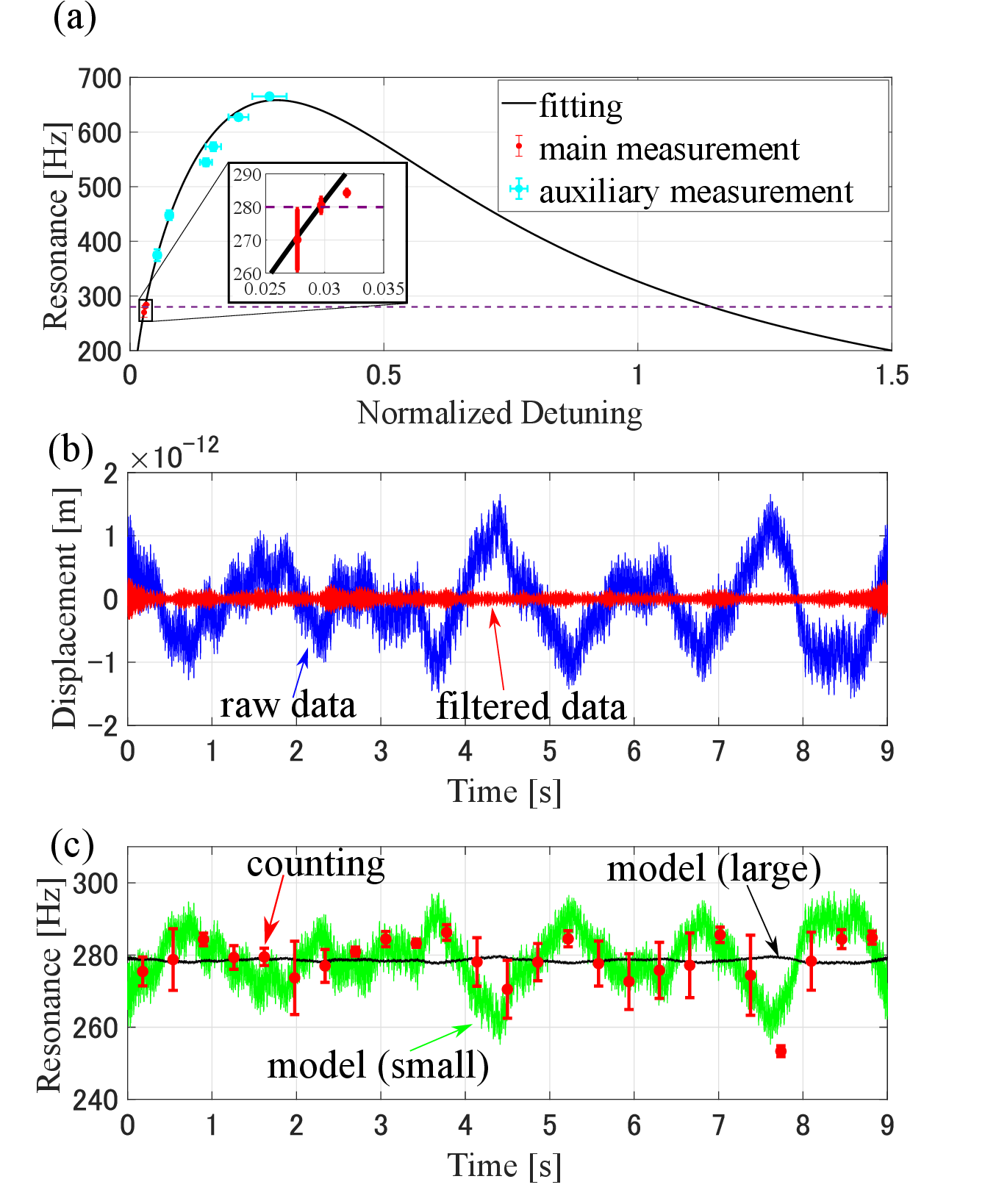}
\centering
\caption{(Color online) The fluctuation of the cavity detuning in the case of the data with the higher mode temperature of 11 mK.
(a) Optomechanical interaction characterized by the optical spring effect. The data of the main measurement (red) is obtained from the red dots in (c). 
(b) The calibrated raw data (blue) and the bandpass filtered data from 170 Hz to 360 Hz (red). 
(c) Variation of the resonance over time. 
The red dots are obtained by the frequency counting from the bandpass filtered data. 
The two curves show the model resonance frequency calculated by 
applying the raw data to the equation of the optical spring for the small 
detuning (green) and the large detuning (black), respectively. 
}
\label{sfig2}
\end{figure}

In order to characterize the optomechanical interaction, we measured the resonance frequency of the optically trapped pendulum with varying detuning. 
The resonance frequency of the optically trapped pendulum is given by
\begin{align} \omega_m=\sqrt{\frac{8\hbar G^2 n_c \delta}{(1+4\delta^2)\kappa m}}. 
%\label{S25}
\end{align}
As explained in the main text, the cavity length was detuned from resonance such that the pendulum's resonance ($\omega_m/2\pi$) increases to 280 Hz. 
Because of the nonlinearity of the optical spring with respect to $\delta$, this leads to a mean value for the detuning of roughly $0.03\times\kappa$ {\it or} $1.2\times\kappa$. 
To determine the detuning based on the optical spring effect, we analyze the variation of the optically trapped pendulum's resonance over time. 
As shown in Fig. \ref{sfig2} (b), the raw data is bandpass filtered around the resonance from 170 Hz to 360 Hz, and then the number of zero crossings is counted in order to estimate its instantaneous resonance frequency. 
Next, the analyzed resonance is low pass filtered with a cutoff frequency of 8.2 Hz, and divided into 25 time bins. 
The result agrees well with the theoretical model for small detuning, as shown in Fig.\ \ref{sfig2} (c). 
The result of the counting is further divided into three bins of different resonance frequency values, and then averaged for each bin. 
Fitting the averaged data with respect to detuning, to the theoretical model shown in Fig.\ \ref{sfig2} (a), the mean value of the detuning is determined to be $\Delta=0.0292(4)\times\kappa$.
Here, we should note that the temperature for the confined mode is relatively high compared to the theoretical prediction given by $T\gamma_0(\omega_{\rm m})/\gamma_{\rm m}$, where $T$ is the room temperature. 
We mainly attribute this to fluctuations of the resonance frequency, namely, the detuning, as shown in Fig.\ \ref{sfig2}. 
As a result, the quantum cooperativity decreases by a factor of $10$.
Furthermore, it also decreases by a factor of $4$ due to mode mixing between the pendulum mode and the dissipative pitching mode \cite{miliG1, sugiyama}. 
Thus, compared to the case without optical spring, the quantum cooperativity of the current experiment is enhanced by a factor of $(\omega_m/\Omega)/10/4=1.5$.

\section{Validity of approximation for verification}
Based on the mean value given by the uncertainty in each optomechanical parameter, the theoretical value for our conditional variances are given by $V_{11}=1242, V_{22}=7850$, and $V_{12}=2386$, respectively. 

Then, we examine the validity of the approximation that can be made between the conditional variances for prediction and retrodiction as given by \eqref{eq:equi}:
\begin{flalign}
\operatorname{Position \hspace{0.1cm}Variance\hspace{0.1cm} Verification}&=V_{11}+V_{E11}\nonumber\\
&=2V_{11}+2\frac{\gamma_m}{\lambda_X}\nonumber%\tag{S27}
\end{flalign}
Since $\gamma_m/\lambda_X=1.27$ is smaller compared to the value of $V_{11}$, we can approximate $V_{E11}+V_{11}\approx 2V_{11}$.
For the momentum verification, this approximation can also be extended:
\begin{flalign}
\operatorname{Momentum \hspace{0.1cm}Variance\hspace{0.1cm} Verification}&=V_{22}+V_{E22}\nonumber%\tag{S28}
\\
&=\frac{2V_{11}+2\gamma_m/\lambda_X}{2\omega^2_m}(2\omega_m(\omega_m+\Lambda_X))+\frac{\gamma_X}{2\omega^2_m} (V^2_{11}+(V_{11}+2\gamma_m/\lambda_X)^2)\lambda_X\nonumber
\end{flalign}
Similarly, we can conclude that $V_{E22}+V_{22}\approx 2 V_{22}$.

Specifically, we have the following values for the relative error in each approximation:
\begin{flalign}
    \frac{(V_{E11}+V_{11})-2V_{11}}{2V_{11}}\times 100\%&=0.1 \% \nonumber\\\frac{(V_{E22}+V_{22})-2V_{22}}{2V_{22}}\times 100\%&=0.2\% \nonumber
\end{flalign}
These values are smaller than the ones we obtained for the modeling error ($\approx 10\%$). Therefore, we can consider our approximation as appropriate for each case.

{We are thankful Seth B. Cata\~{n}o-Lopez for help with the manuscript and discussions. 
We thank Keiichi Edamatsu and Daisuke Miki for discussions. 
We thank Chao Meng for answering our questions. 
This research is supported by JSPS KAKENHI Grant No. 15617498 and JST FORESTO Grant No. JPMJFR202X.}
%JST CREST Grant Number JPMJCR1873,
%, and JST PRESTO Grant No. JPMJPR166A.}

% Bibliography
%\bibliography{pnas-sample}

\end{document}